# EFFECTIVE INTEGRATION OF ICT TO FACILITATE THE SECONDARY EDUCATION IN SRI LANKA

*Palagolla W W N C K and Wickramarachchi A P R*

## ABSTRACT

Information and Communication Technology (ICT) has been the phenomenon of the 21$^{st}$ century and its growing advances continue to play a vital role as an ideal tool to acquire, store, disseminate and apply knowledge than ever before. Thus, the integration of ICT in diverse business processes has increased the importance as an imperative source of economic growth in the rapidly changing knowledge economy. The paper mainly explores potential barriers towards the effective integration of ICT and its impact on the performance of the secondary education. A structured survey questionnaire gathered empirical data from a random sample of teachers from selected schools in the North Central Province (NCP) of Sri Lanka. Results show very low integration of ICT in schools and teachers' ICT competency. ICT infrastructure, leadership support and school planning are revealed as major organizational constraints for the effective integration of ICT in schools. In contrast, respondents' fairly positive attitudes towards ICT indicate potential future developments. Comparative findings revealed that ICT education and English language proficiency are significant demographic predictors of ICT utilization. Results also reported a positive impact of ICT on teachers' job performance.

## INTRODUCTION

Information and Communication Technology (ICT) has become one of the basic building blocks of the modern society. The radical technological transformation in both developed and developing countries has made pervasive impacts on various segments. Therefore, it is not surprising to find an exponential growth in the use of ICT in education all over the world. Some impressive evidence on the effectiveness of ICT in education reveals that it has greater impact than any other innovation (Fluck, 2003). The emergence of the knowledge economy has also brought much greater emphasis on education. Almost all countries now regard ICT competency as a part of the core of education that facilitates to develop students' capacities for self-learning, problem solving, critical thinking and collaboration etc (Yuen, Law, & Wong, 2003). Although the emerging technology based solutions have become an integral part of teaching, learning, and school management, considerable number of barriers have made this ineffective in most countries. In Sri Lanka, the problem is more prominent among schools in the rural areas than the urban areas. For instance, NCP is geographically the largest province of the country which is one of the areas that suffer with this problem.

The main objective of this paper is to explore potential barriers towards the effective integration of ICT and its impact on job performance in secondary education. It also recommends school level changes as well as policy level changes in order to address identified issues. Therefore, the findings would be of greater significance for the policy and decision makers, school administrators and stakeholders in the secondary education system of Sri Lanka. Furthermore, the findings would help to address the dearth of empirical investigations of this field of study. It is also presumed that investigating ICT usage and integration into schools is crucial at present as the government has declared 2009 as the year of ICT and English.

## LITERATURE REVIEW

The integration of ICT in the secondary education is a national policy across many countries, which was initiated in Sri Lanka in 1980s. There is enough evidence to suggest that ICT has the potential to impact on every aspect of the school activities. Thus, schools cannot remain as mere venues in the fast growing technological transformation. They must promote an effective use of ICT in order to promote new ways of teaching and learning, information management, professional development, creativity etc. Developing a set of ICT based practices to capture new knowledge, configure and store them in various formats, disseminate them in effective ways for quick grasping, and apply knowledge in more innovative ways would improve services and outcomes of schools in diverse aspects. Specially, it would help students to reach their full potential. The following sub sections review the use and impact of ICT, and barriers towards integrating ICT in schools. Further, it also discusses the present status of ICT use in secondary education in Sri Lanka.

## USE & IMPACT OF ICT IN EDUCATION

Information and Communication Technology (ICT) is defined as a diverse set of technological tools and resources used to communicate, and to create, store, disseminate, and manage information. The technologies include broadcasting technologies (radio and television) as well as newer digital technologies such as computers and the Internet, which enable set of powerful tools for educational change and reform. Eadie (2000) found that most schools in Australia, USA, England, and Hong Kong have integrated innovative ICT tools to support school practices. Technologies like shared software, videoconferencing, digital imaging and editing facilities, video walls for image projection, and online-learning communities are used in schools that allow creating and disseminating knowledge more effectively. Further, chat and instant messaging, virtual art gallery, and virtual museum are tremendous information sharing technologies used in schools. Livetext is another new technology that allows teachers to put up content on a web page and enables online classes. Virtual learning systems are useful tools to store information digitally. Interactive whiteboards transform traditional black boards into an entirely different interactive teaching tool. Condie and Munro (2007) describes mobile technologies, learning platforms, and virtual learning environments as information dissemination technologies, which are fast becoming central to whole range of tools that support school activities. Further, students with special educational needs are also supported with specialist technologies such as speech recognition software and specialist peripherals. Learning platforms and e-portfolios provide a range of ICT based functions around communication and collaboration, content management, curriculum planning. E-portfolios are larger personal online spaces that allow users to store, organize, and personalize information, collaborate and receive feedback (Becta, 2005). In addition, diverse tools and services such as email, messaging, discussion forums, and blogs could also play a significant role. The effective use of ICT in schools continues to rise steadily. ICT is now widely recognized as an essential tool for teaching and learning in the 21st century. It is noticeable that most teachers regard ICT positively and report increased use of computers for planning, preparing presentations, worksheets and other learning materials, administration, assessment and tracking students' progress (Ofsted, 2004). The effective integration of ICT in education is a complex and multifaceted process. The appropriate use of ICT expands access to education, strengthens the relevance of education to highly digital work environments, and raises educational quality (Tinio, 2003). Kimble (1999) shows that technology can result in increased student self-confidence and eagerness to learn. Balanskat, Blamire, & Kefala (2006) presents that ICT can impact positively on students' educational performance, motivation, attention, collaboration, and communication and process skills. On the other hand, it shows considerable evidence regarding the impact of ICT on teachers' increased enthusiasm, efficiency, and collaboration. Newhouse (2002) reported positive impacts of ICT on curriculum, pedagogy, students' learning, and learning environments. It also provided evidences on improvements in active learning, productivity, motivation, higher level thinking, independence, collaboration, and overcoming physical disabilities with the effective integration of ICT in the classrooms. Newhouse, Trinidad, & Clarkson (2002) also noted an effective integration of ICT in the classrooms enables teachers to adopt a balanced pedagogical approach between teacher-centered instruction and learner-centered collaborative environment. Bailey, Day, Day, Griffin, Howlett, Kane, Kirk, McCullough, McKiernan, McMullen, Perfect, Ramsey, & Wood (2004) indicated that ICT had enabled teachers to become more efficient with the better management, storage, and maintenance of work. It is also important to note that ICT plays a very effective role in the higher education too. As shown in Kumar and Kumar (2006), the benefits of ICT based practices in the Indian higher education are: improved quality in research and development, curriculum development, administration, students' affairs, and planning and development. However, previous literature also indicated the negative impact of ICT in schools. As noted by Kimble (1999), most negative results are in line with the way technology is used in the classroom, the technical expertise and preparedness of teachers, and the relative cost of acquiring the technology.

## BARRIERS FOR THE EFFECTIVE USE OF ICT IN SCHOOLS

Although the use of ICT in education has been a priority in most countries from the last decade, considerable barriers still exist. Some schools in some countries have integrated ICT into the curriculum and have transformed teaching and learning with the use of innovative technologies. However, most schools across the world are still in the early stage in adopting ICT and no records for significant improvements due to considerable barriers (Becta, 2005). Therefore, in order to make realistic and holistic solutions for the issues, factors that prevent the full use of ICT in schools must be clearly identified. Balanskat et al. (2006) has divided the perceived barriers in schools into three broader categories: teacher level barriers, school level barriers, and system level barriers. The teacher level barriers incorporate factors related to teachers' attitudes and approach to ICT such as lack of ICT skills, lack of motivation and confidence on ICT, and inappropriate teacher training. School level barriers include those related to the institutional context such as the absence and/or poor quality of ICT infrastructure, limited access to ICT equipment, school's limited project-related experience, lack of experience in project-based learning, and absence of ICT mainstreaming into schools' strategies. The system level barriers are those related to the wider educational framework which mainly focuses on the rigid structure of the traditional schooling system. It is commonly accepted that the effective use of ICT requires more than just the technology and competent teachers. Newhouse (2002) pointed out essential conditions for the effective implementation of ICT. Some of the most significant conditions are: proactive leadership, technical assistance, financial support, culture, policies and procedures, training and support, and provision of hardware and software infrastructure.

## ICT IN SECONDARY EDUCATION OF SRI LANKA

Computer based education in secondary schools in Sri Lanka was first started in 1983. At present, it has improved considerably. However, the progress is uneven within the country. The Ministry of Education (MoE) of Sri Lanka is also of the view that Sri Lanka is still lagging behind in integrating ICT when compared to other countries with similar economic and social status. Therefore, in an effort to fully utilize and improve ICT in secondary education, a national policy on school IT education was approved in 2001. As a result, MoE introduced ICT as a subject at the GCE O/L in 2001 (MoE, 2007). Meanwhile, several foreign funded projects supported to improve ICT usage in Sri Lankan schools. In 2001-2004, the World Bank funded General Education Project II (GEP II) established 400 ICT Centers in schools. The Asian Development Bank (ADB) funded Secondary Education Modernization Project (SEMP) established 1006 Computer Learning Centers (CLCs) by 2006 and at present SEMP II expanding activities further. Up to one million secondary students will be benefited by these programs to modernize the secondary education in Sri Lanka (ADB, 2008; MoE, 2007). To enhance teaching and learning through ICT, several initiatives were taken place in secondary schools in Sri Lanka. The major initiatives include a dedicated educational Web portal, Internet access for more than 1000 schools and 120 educational institutions via SchoolNet service, provision of multimedia rooms in order to use learning materials in key subjects, development of MoE Website in order to facilitate e-learning and information management, establishment of hardware and network solution teams at school level, and the implementation of numerous annual educational software competitions to inculcate ICT skills in the Sri Lankan school system.

## KEY CHALLENGES

In the new Millennium, computer literacy of a nation will be one of the major determinants to measure the level of education. However, successful integration of ICT in Sri Lankan schools is challenged by number of major issues including:

- Lack of teacher training programs on the effective integration of ICT in teaching and learning.
- Poor English language proficiency among school teachers that mainly hinder the use of World Wide Web.
- Lack of connectivity especially in remote, rural, and disadvantaged areas.
- Lack of teacher guides, resource books, and model question papers to support the curriculum.
- Lack of awareness and resistance due to poor attitudes and motivation towards ICT among teachers and principals.

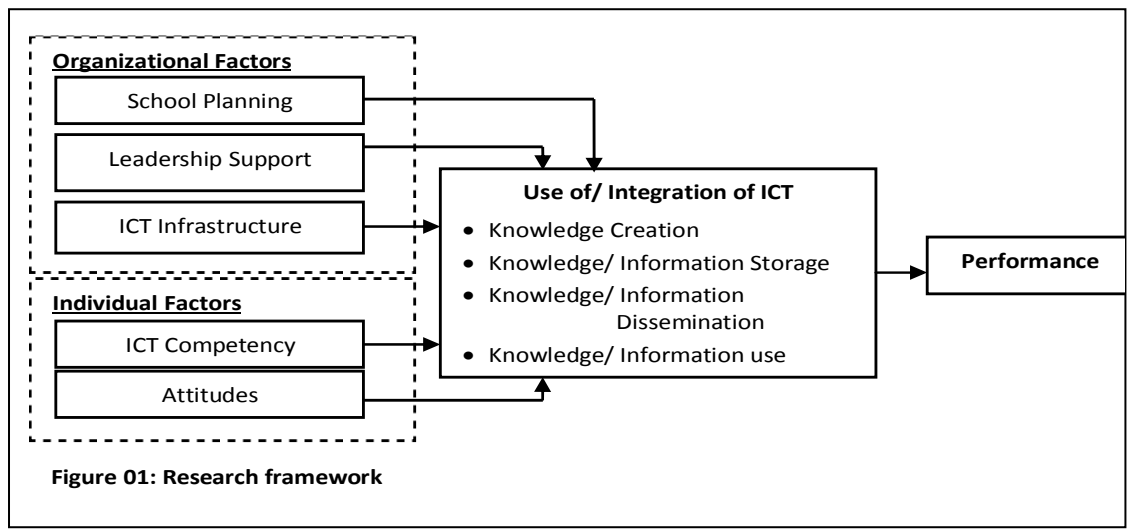

Figure 01: Research framework

## METHODOLOGY

A survey was conducted to explore potential barriers that impede effective integration of ICT in schools. A total of 145 teachers were drawn randomly from 30 ICT facilitated secondary schools in the NCP of Sri Lanka. A structured questionnaire was used for collecting empirical data.

## RESEARCH FRAMEWORK

A research framework was developed based on similar studies (Becta, 2007; Li, 2007; Muhammed, 2006) conducted previously (see Figure 01). The framework identifies two categories of factors that affect the integration of ICT in schools, namely: organization factors and individual factors. School planning, leadership support and ICT infrastructure were part of organizational factors. Individual factors included ICT competency and attitudes of individuals towards use of ICT. Ultimately these factors affect the performance of both individuals as well as the school as a whole.

In order to identify factors that affect effective integration of ICT, six hypotheses as given below were developed.

*H1: Use of ICT is very low among teachers of the ICT facilitated secondary schools in the NCP of Sri Lanka.*

*H2: Attitudes of individuals towards ICT is the most influencing factor for the effective use of ICT.*

*H3: The level of ICT use is higher among staff with ICT education than staff without ICT education*

*H4: The level of ICT use is proportionate to the level of English language proficiency among staff.*

*H5: The level of ICT use is higher among staff with ICT training than the staff without ICT training.*

*H6: There is a strong positive relationship between the use of ICT and job performance.*

## FINDINGS

Survey responses were rated on a five point Likert type scale ranging from "Very Low (1)" to "Very High (5)". The statistical tools employed in the data analysis are descriptive statistics (Mean score and Frequencies), Pearson's correlations, and the Analysis of Variance (ANOVA).

## ICT USE IN THE SCHOOLS IN NCP

According to the mean (M) score analysis, the level of ICT integration revealed very low (M = 1.99) among school teachers in the NCP (see Table 01). Further, the analysis of frequencies show only a minority of teachers (15.9 %) integrate ICT to a higher level for their day to day activities while the majority (84.1%) report with a low to moderate integration. Therefore, the finding supports *H1*.

A reason for this observation could be lack of ICT qualified human resources as NCP is one of the backward provinces of Sri Lanka. It could also be concluded some of the government initiatives such as new graduate teacher appointments, increased ICT training opportunities, establishment of ICT units at school level, and the provision of Computer Resource Centers (CRCs) on regional basis to implement ICT are still not yielded positive results.

**Table 01: Use of ICT in schools in NCP**

Scale: Low (M < 2.5), Moderate (M < 3.5), and High (M ≥ 3.5)

**BARRIERS FOR EFFECTIVE**

**INTEGRATION OF ICT**

The paper presents several individual level and organizational level factors that contribute towards poor utilization of ICT in secondary schools of the region (see Table 02). Pearson correlation analysis was used to identify important barriers. Strong positive correlations were found for all five factors investigated. Accordingly, ICT infrastructure, leadership support, school planning, ICT competency, and attitudes respectively are the major factors that hinder the effective integration of ICT in the secondary education. One optimistic point is that individual attitudes are fairly positive with regard to the ICT integration with a mean score 3.10. Thus, the hypothesis *H2* could be rejected.

**Table 02: Barriers for effective use of ICT**

| Barriers for ICT use | Correlation | Mean |
|---|---|---|
| ICT Infrastructure | 0.82** | 2.09 |
| Leadership Support | 0.80** | 2.04 |
| School Planning | 0.78** | 2.03 |
| ICT Competency | 0.75** | 2.79 |
| Attitudes towards ICT | 0.70** | 3.10 |

Scale: Low (M < 2.5), Moderate (M < 3.5), and High (M ≥ 3.5)
Correlation: Weak (r < 0.5), Moderate (r = 0.5), Strong (r > 0.5)
Correlations are significant at: *p< 0.05, **p< 0.01

However, there is still a considerable dearth of ICT infrastructure especially in the rural Sri Lanka. The paper highlights the inadequate IT resources (r = 0.82) such as computer hardware, software, and ICT based training opportunities among secondary schools in the NCP. Therefore, it is prudent to implement both government and organizational interventions in order to address the issue. Furthermore, the open ended responses of the survey indicate other potential barriers towards the effective integration of ICT in secondary schools in the NCP including lack of time, higher workload, inadequate financial resources, lack of public institutions to learn ICT, and fear towards new technologies.

**INDIVIDUAL DIFFERENCES IN ICT USE**

The paper underlines significant comparative

| ICT use | Mean | Staff Percentages | | |
|---|---|---|---|---|
| | | Low | Moderate | High |
| Overall Use | 1.99 | 71.7 | 12.4 | 15.9 |
| Knowledge/ Information Creation | 2.05 | 69.7 | 13.7 | 16.6 |
| Knowledge/ Information Storing | 1.92 | 74.5 | 09.6 | 15.9 |
| Knowledge/ Information Disseminating | 2.02 | 71.0 | 12.4 | 16.6 |
| Knowledge/ Information Use | 1.98 | 73.1 | 11.7 | 15.2 |

information regarding the use of ICT among sub groups of individuals (see Table 03).

**Table 03: Individual differences in ICT use**

| Sub Group of Individuals | Mean |
|---|---|
| **ICT Education**: Yes (N=69) | 2.34 |
| No (N=76) | 1.68 |
| **English Language Proficiency**: | |
| High (N=33) | 2.60 |
| Medium (N=56) | 2.09 |
| Low (N=56) | 1.54 |
| **ICT Training**: Yes (N=63) | 2.00 |
| No (N=82) | 1.99 |

Scale: Low (M < 2.5), Moderate (M < 3.5), and High (M ≥ 3.5)

As shown in Table 03, ICT utilization is relatively high among staff with ICT education than staff without ICT education. Thus, *H3* is supported by the study. Similarly, it was found that the use of ICT is proportionate to the English language proficiency. It means that the English language proficiency plays a vital role in ICT use that supports *H4*. However, it is worth noting that only a slight difference in use of ICT between staff with and without ICT training. This may be due to ineffectiveness of the current training programs. Respondents have commented that the existing training programs provide very limited support on how to integrate ICT at work. Therefore, well designed training programs on task-technology fit may motivate and encourage individuals to use ICT at work.

Table 04 analyses differences in ICT competency and attitudes towards ICT among staff with and without ICT education, English language proficiency, and ICT training provided. As shown in the findings, adequate ICT education may provide competency through confidence, and motivation for creative integration of ICT into teaching, learning, and related work. It has also contributed in developing positive attitudes towards ICT.

It could also be noted that English language proficiency also plays an important role in improving ICT competency and positive attitudes towards ICT. As many books on ICT are available in English and the standard language used in computers is English in Sri Lanka, English proficiency may play a significant role.

**Table 04: Individual differences in ICT competency and attitudes towards ICT**

| Sub Group of Individuals | ICT Competency | Attitudes towards ICT |
|---|---|---|
| **ICT Education:** | | |
| Yes | 3.42 | 3.52 |
| No | 2.21 | 2.72 |
| **English Language Proficiency:** | | |
| High | | |
| Medium | 3.60 | 3.63 |
| Low | 3.06 | 3.24 |
| | 2.04 | 2.65 |
| **ICT Training:** | | |
| Yes | 2.86 | 3.12 |
| No | 2.73 | 3.08 |

Scale: Low (M < 2.5), Moderate (M < 3.5), and High (M ≥ 3.5)

Results also indicate that majority of staff has not received any ICT training. However, they also show that ICT training has not much contributed in improving ICT competency as well as attitudes of individuals towards ICT. Quality of ICT training programs conducted and target audience selected for training may have played a part in this. All things considered, it can be concluded that *H5* is slightly supported by the findings.

**THE USE OF ICT AND JOB**

**PERFORMANCE OF STAFF**

Impact of ICT usage on job performance is examined in Table 05. It shows positive relationships between all dimensions and performance at higher levels of significance. It is notable that higher the use of ICT, higher the job performance and vice versa. However, all these relationships were shown as moderate. This may be mainly due to lack of ICT infrastructure, poor leadership support, inadequate school planning, and lack of ICT competency. All these factors and attitudes towards ICT are generally known as fundamental prerequisites for the effective use of ICT in any environment.

**Table 05: Relationship between ICT use and job performance**

| ICT use | Performance |
|---|---|
| Overall use | + 0.51** |
| Knowledge/ information creation | + 0.49** |
| Knowledge/ information storage | + 0.51** |
| Knowledge/ information dissemination | + 0.50** |
| Knowledge/ information use | + 0.51** |

Scale: Weak (r < 0.5), Moderate (r = 0.5), Strong (r > 0.5)
Correlations are significant at: *p< 0.05, **p< 0.01

**CONCLUSION**

This research presents the current status of ICT usage in secondary schools in NCP of Sri Lanka. It was identified that lack of ICT infrastructure, poor leadership support, inadequate school planning, lack of ICT competency, and attitudes towards ICT respectively as major barriers for the effective integration of ICT in schools. It also highlights the role played by ICT education, English language proficiency and ICT training provided to staff. ICT's impact on performance of staff found to be positive. In order to integrate ICT into schools effectively, following policy level as well as organizational level solutions could be recommended. Following government interventions are vital to improve the use of ICT in schools.

Increasing financial allocations for a better supplement of ICT infrastructure.
Increasing the number of ICT training centers at the provincial level and organizing effective ICT training programs.
Improving English language skills.

Focusing special attention on improving leadership capabilities among school principals and ICT centre managers/coordinators.

Planning and setting targets for the effective integration of ICT in schools

Introducing a benchmarking system to identify best practices.

Although policy level changes are crucial, the importance of organizational level changes could not be undermined. However, these changes have to be designed and implemented in line with the policy level initiatives.

Organizing workshops, seminars, and different programs to build positive attitudes towards ICT.

Providing opportunities for all staff to participate ICT training programs that offer at the national level.

Providing an appropriate level of support and direction for staff and students.

Proper planning for the use and improvement of ICT.

Establishing an ICT committee for each school to support, motivate, and assess the use of ICT in schools.

Introducing a periodic review and reward system to promote the use of ICT at the schools.

Providing opportunities and encouraging staff as well as students to use ICT for their daily activities.

Due to time and resource constraints, this study was restricted to the NCP of Sri Lanka. Thus, the findings may not be able to generalize to the entire country as NCP is relatively an underdeveloped area. Further, the research framework can be improved by including more aspects that affect the use of ICT in secondary schools. Since this is an important attempt to investigate the integration of ICT in secondary schools of the NCP, longitudinal replications at national level will help to validate the issues identified.